\documentclass{emulateapj}
\usepackage{lineno}

\usepackage{amsmath}
\usepackage{epsfig}
\usepackage{amssymb}
\usepackage{graphicx}
\usepackage{color}
\usepackage{natbib}
\usepackage{soul}

\usepackage{apjfonts} 
\usepackage{amsmath,amstext}
\usepackage[breaklinks,colorlinks,citecolor=blue,linkcolor=magenta]{hyperref} 
\usepackage[all]{hypcap} 
\usepackage{verbatim}

\begin{document}

\title{A Simple Condition for Sustained Super-Eddington Black Hole Growth}

\author{Jarrett Lawrence Johnson\altaffilmark{1} and Phoebe R. Upton Sanderbeck\altaffilmark{2}}

\altaffiltext{1}{XTD-DO, Los Alamos National Laboratory, Los Alamos, NM 87545}
\altaffiltext{2}{XTD-IDA, Los Alamos National Laboratory, Los Alamos, NM 87545}

\begin{abstract}
One of the most pressing questions in cosmology is how the black holes (BHs) powering quasars at high redshift grow to supermassive scales within a billion years of the Big Bang.  Here we show that sustained super-Eddington accretion can be achieved for BHs with Eddington fractions $f_{\rm Edd}$ $\ga$ 2/$\epsilon$, where $\epsilon$ is the efficiency with which radiation is generated in the accretion process.  In this regime, the radiation carries too little momentum to halt the accretion flow and the infalling gas traps the radiation.  The BH growth then proceeds unimpeded until the gas supply is exhausted, in contrast to accretion at lower rates which is limited by the radiation generated in the accretion process.  The large gas supply available in massive high-redshift quasar host galaxies may be readily accreted onto seed BHs via this supply-limited mode of accretion, providing an explanation for how such supermassive BHs are assembled in the early universe.  This sustained super-Eddington growth may also explain the short lifetimes inferred for the H~II regions surrounding high-redshift quasars, if the bulk of the BH growth occurs without the associated radiation escaping to ionize the intergalactic medium.  It furthermore implies that a population of obscured rapidly growing BHs may be difficult to detect, perhaps explaining why so few quasars with Eddington fractions higher than a few have been observed.  Finally, this simple condition for sustained super-Eddington growth can easily be implemented in cosmological simulations which can be used to assess in which environments it occurs.
\end{abstract}


\maketitle

\section{Introduction}
There remain key open questions about the growth of the earliest black holes (BHs), in particular with regard to their progenitors and to how they grow rapidly enough to power bright quasars existing within the first billion years after the Big Bang, a large and growing number of which have been detected \citep{mortlock11,wu15,banados18,yang20,bosman20}. In order to explain their prompt emergence, it is clear that they must either grow from very massive so-called seed BHs (e.g. \citealt{shang10,volonteri12,Smith2017,Woods2019}) and/or grow at average rates that are extraordinarily high, perhaps even in excess of the Eddington rate at least intermittently (e.g. \citealt{tanaka09,alexander2014,madau14,volonteri15,lupi2016,pezzulli17a, massonneau2022}). 

Over the past decade, analytical models of accretion onto BHs at super-Eddington rates have explored the conditions under which such flows can be established and facilitate the accelerated accretion of large supplies of gas in early galaxies (e.g. \citealt{wyithe12,pacucci15,sakurai16,levinson18,takeo2019}).  Accompanying 
these are more detailed multi-dimensional simulations that confirm the key physics at play in these flows, in particular photon trapping \citep{begelman78}, that allow for super-Eddington accretion to be sustained (e.g. \citealt{oshuga05,jiang2014,sadowski16}).  While this work collectively supports the possibility of early supermassive BH growth by super-Eddington accretion, it remains for this theory to be tested decisively against observations (e.g. \citealt{pognan2020}) and it has only recently been incorporated into large-scale cosmological simulations (e.g. \citealt{mayer2019,scoggins2022}).  

Here we derive a simple analytical condition for gas supply-limited super-Eddington accretion, presented in Section~\ref{sec:derivation},  that lends itself to comparison with data, as well as to adoption in cosmological simulations of quasar formation.  We compare our derivation to previous theoretical work in Section~\ref{sec:comparison} and then make the comparison to supermassive BH data in Section~\ref{sec:data}.  Finally, we discuss the implications of our results in Section~\ref{sec:implications}.


\section{Derivation of the condition}
\label{sec:derivation}
We begin by assuming that the two main forces that dictate the
nature of the accretion flow onto a BH are gravity and that from the
radiation emitted in the accretion process, which we assume to be produced deep in the accretion flow near the BH. We furthermore assume a steady-state accretion flow. In Section~\ref{sec:spherically_symm}, we assume a spherically symmetric flow, and in Section~\ref{sec:accretion_disk} we consider  flow through an accretion disk.  

\subsection{Spherically Symmetric Accretion Flow}
\label{sec:spherically_symm}
We assume that
the gravity of the BH dominates at large distances $r$, such that the
infall velocity of the gas is well-approximated as

\begin{equation}
\label{eq:infall_vel}
v = \, \left(\frac{2 \, G \, M_{\rm BH}}{{r}}   \right)^{\frac{1}{2}} \mbox{\ ,}
\end{equation}
where $M_{\rm BH}$ is the mass of the BH and $G$ is the
gravitational constant.

Under the assumption of steady-state spherically symmetric accretion at a rate ${\dot
  M}_{\rm BH}$, in which case ${\dot
  M}_{\rm BH}$ = 4$\pi$$r^2$$\rho$$v$, we can express the density $\rho$ of the gas as

\begin{equation}
\label{eq:gas_dens}
\rho = \frac{1}{4 \pi \left(2 G M_{\rm BH}   \right)^{\frac{1}{2}}}  \frac{{\dot M}_{\rm BH}}{r^{\frac{3}{2}}} \mbox{\ .}
\end{equation}

The rate at which momentum passes through the spherical surface at radius $r$ from the BH, purely under the influence of the gravity of the BH, is given by 

\begin{equation}
\label{eq:pgrav}
{\dot p}_{\rm grav} = {\dot M}_{\rm BH} v(r) \mbox{\ .}
\end{equation}

In turn, the maximum rate at which momentum can be carried outward through the same surface by the radiation generated in the accretion process is given by 

\begin{equation}
\label{eq:prad}
{\dot p}_{\rm rad} = \epsilon {\dot M}_{\rm BH} c \mbox{\ ,}
\end{equation}
where $c$ is the speed of light, $\epsilon$ is the efficiency with
which radiation is generated as gas is accreted onto the BH, and we have utilized the fact that the momentum $p_{\rm rad}$ and energy $E_{\rm rad}$ carried by radiation are related by $p_{\rm rad}$ = $E_{\rm rad}$/$c$. Equating ${\dot p}_{\rm grav}$ to ${\dot
  p}_{\rm
  rad}$, and inserting Equation~\ref{eq:infall_vel}, yields the following expression
for the distance from the BH within which the momentum in the accretion flow exceeds that of the radiation emitted from the gas accreting onto the BH, implying that the accretion flow will not be halted by the radiation:

\begin{equation}
  r_{\rm eq} = \frac{2 G M_{\rm BH}}{\epsilon^2 c^2} 
                 = R_{\rm S} \epsilon^{-2} \mbox{\ ,}
               \end{equation}
where R$_{\rm S}$ is the Schwarzschild radius.

This condition holds only so long as the infalling gas outside of
$r_{\rm eq}$ is not impacted by the
radiation emitted near the BH, as we have assumed so far.  In order to
check this assumption, we first calculate the mean free path $\lambda$
for photons at $r_{\rm eq}$:

\begin{equation}
\lambda \simeq \frac{1}{\sigma_{\rm T} n_{\rm e}}  \mbox{\ ,}
\end{equation}  
where $\sigma_{\rm T}$ is the Thomson cross section and we have
assumed a fully ionized gas for which the number density of electrons
is $n_{\rm e}$ $\simeq$ $\rho$ / $m_{\rm H}$, with $m_{\rm H}$ the
mass of the hydrogen atom.  Using the expression for $\rho$ in
Equation~\ref{eq:gas_dens} above, we then have

\begin{equation}
\lambda = \frac{4 \pi m_{\rm H}}{\sigma_{\rm T} {\dot M}_{\rm BH}}
\frac{R_{\rm S}^2 c}{\epsilon^3} \mbox{\ }
\end{equation}
at $r_{\rm eq}$.
With this expression, the timescale on which the radiation emitted
near the black hole diffuses outward at $r_{\rm eq}$ can be estimated
as

\begin{equation}
\label{eq:tdiff1}
t_{\rm diff} \simeq \frac{r_{\rm eq}^2}{c \lambda} = \frac{{\dot
    M}_{\rm BH} \sigma_{\rm T}}{4 \pi
  m_{\rm H} c^2 \epsilon} \mbox{\ .}
\end{equation}

\begin{figure}
   \includegraphics[angle=90,width=3.8in]{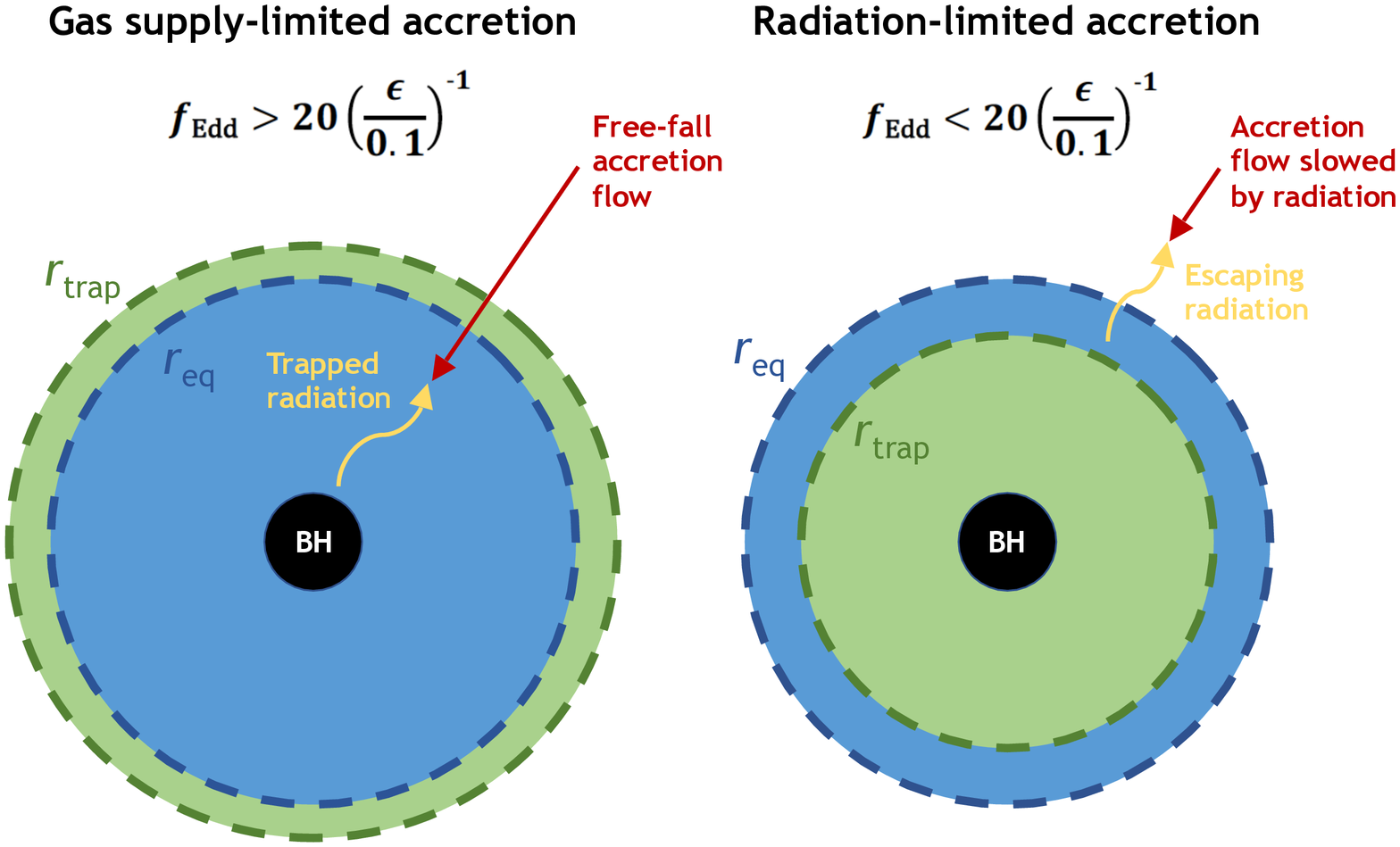}
   \caption{Comparison of accretion flows at rates above ({\it left}) and below ({\it right}) the critical rate $f_{\rm Edd,crit}$ for sustained super-Eddington accretion given by Equation~\ref{eq:f_edd_crit}.  At rates below $f_{\rm Edd,crit}$, radiation produced within $r_{\rm eq}$ is able to escape beyond $r_{\rm eq}$ and limit the accretion of gas onto the BH.  At rates above $f_{\rm Edd,crit}$, this radiation is trapped in the flow and the accretion of gas and radiation onto the BH proceeds unimpeded.}
   \label{fig:accretion_compare}
   \end{figure}

If this timescale is longer than the timescale $t_{\rm ff}$ $\simeq$
$r_{\rm eq}$ / $v$ for the gas at $r_{\rm eq}$ to fall toward the BH,
then the radiation will be trapped within $r_{\rm eq}$, as shown in Figure~\ref{fig:accretion_compare}.  Consequently, the gas will arrive at $r_{\rm eq}$ with more
momentum than the radiation within $r_{\rm eq}$ and it  will thus be accreted onto the BH.

Taking the ratio of $t_{\rm diff}$ / $t_{\rm ff}$ at $r_{\rm eq}$ and
again inserting Equation~\ref{eq:infall_vel} for $v$, we have

\begin{equation}
\label{eq:t_ratio}
\frac{t_{\rm diff}}{t_{\rm ff}} = \frac{\sigma_{\rm T} c \epsilon^2}{8
    \pi m_{\rm H} G} \frac{{\dot M}_{\rm BH}}{M_{\rm BH}} \mbox{\ .}
\end{equation}


Setting this ratio equal to one and normalizing to the Eddington
accretion rate ${\dot M}_{\rm BH, Edd}$ = $L_{\rm Edd}$~/~($\epsilon$$c^2$), where $L_{\rm Edd}$ is the Eddington luminosity, yields 

\begin{equation}
\label{eq:f_edd_crit}
 f_{\rm Edd, crit} \simeq 20 \, \left(\frac{\epsilon}{{0.1}}   \right)^{-1}
 \mbox{\ ,} 
\end{equation}
where $f_{\rm Edd, crit}$ is the fraction of the Eddington accretion rate
above which sustained super-Eddington
accretion proceeds unimpeded by the radiation emitted near the BH.\footnote{For clarity, we note that the key difference between this derivation and the force balance in the derivation of the Eddington limit is that in this derivation the gravitational force is the only force acting at large radii ($r$ > $r_{\rm eq}$), due to the radiation being trapped at small radii ($r$ < $r_{\rm eq}$).}
As long as the accretion rate exceeds $f_{\rm Edd,crit}$, and under our assumptions of
spherical symmetry and a steady-state flow, accretion onto the BH will continue until the gas supply is exhausted.

\subsection{Flow Through an Accretion Disk}
\label{sec:accretion_disk}
While the result above has been derived under the assumption of a spherical accretion flow, it is important to consider the impact that accretion through a disk would have. One of the key differences would be that the radial infall velocity through a disk is lower than the free-fall velocity given by Equation~\ref{eq:infall_vel}, as material with non-zero angular momentum about the BH piles up in the disk before finally being accreted onto the BH. In turn, because ${\dot M}_{\rm BH}$ $\propto$ $r^2\rho v$ under the assumption of steady-state accretion, at a given distance $r$ from the BH, and in particular at $r_{\rm eq}$, this drop in radial infall velocity $v$ accompanies the increase in density $\rho$ $\propto$ $v^{-1}$ of the gas as it piles up in the disk.  Considering the two timescales that appear in Equation~\ref{eq:t_ratio}, the drop in infall velocity implies an increase in $t_{\rm ff}$ $\propto$ $v^{-1}$, while the increase in density implies a corresponding increase in $t_{\rm diff}$ because $\lambda$ $\propto$ $\rho^{-1}$.  Thus, the ratio of these timescales $t_{\rm diff}$~/~$t_{\rm ff}$ remains unchanged and Equation~\ref{eq:f_edd_crit} remains valid when accounting for the pile-up of gas in the disk and the corresponding slowing of the radial infall velocity, under the steady-state assumption.  

Additionally, the flow is compressed within the disk such that it passes through an area smaller than 4$\pi r^2$ at a given distance $r$.  In turn, because ${\dot M}_{\rm BH} = 4\pi r^2 \rho v$ under the steady-state assumption, at a given accretion rate this implies an increase in the product $v \rho$ at a given distance $r$, and in particular at $r_{\rm eq}$, due to disk formation.  Because $t_{\rm ff}$ $\propto$ $v^{-1}$ and $t_{\rm diff}$ $\propto$ $\rho$, the effect of the constriction of the accretion flow through the disk is to increase the ratio of these timescales relative to Equation~\ref{eq:t_ratio}.  In turn, we expect that the main effect of disk formation is to lower $f_{\rm Edd,crit}$ relative to the value given in Equation~\ref{eq:f_edd_crit}, which we thus consider an upper limit for the critical steady-state accretion rate above which radiation cannot halt the infall of gas onto the BH. 


\section{Comparison to previous theory}
\label{sec:comparison}
Here we briefly discuss how our results compare to previous analytical and simulation results.

Our definition of $t_{\rm diff}$ is related to the trapping radius $r_{\rm trap}$ presented in \citet{begelman78} by $t_{\rm diff}$ = $r_{\rm trap}$~/~($\epsilon c$). Noting from the balance of Equations~\ref{eq:pgrav} and~\ref{eq:prad} that the free-fall velocity of the gas is $v$ = $\epsilon c$ at $r_{\rm eq}$, this implies that, in essence, we have solved for the condition under which $r_{\rm trap}$ $>$ $r_{\rm eq}$, as shown schematically in Figure~\ref{fig:accretion_compare}. In this, our assumption is that the vast majority of the radiation is generated deep in the accretion flow near the BH, such that we can neglect the small fraction produced outside $r_{\rm eq}$.  This should be a valid assumption as, for example, for a case with a canonical value of $\epsilon$ = 0.1, the gravitational potential energy of the accreting material $E_{\rm grav}$ $\propto$ $G$$M_{\rm BH}$~/~$r$ that is available for conversion to radiation outside $r_{\rm eq}$ = 100~$R_{\rm S}$ is only of order one percent that which is available deep in the accretion flow near the BH at $r$ $\ga$ $R_{\rm S}$. 
Indeed, full simulations clearly show that the hottest, most strongly emitting gas is found to be produced well within $r_{\rm eq}$ (e.g. \citealt{kawaguchi03,watarai06}).  

Another critical assumption in our derivation is that, even within $r_{\rm trap}$, momentum is transferred from the radiation generated in the flow to the infalling gas, through Thomson scattering.  In fact, this must be the case, as Thomson scattering is the very process by which the gas traps the radiation within $r_{\rm trap}$.  Failing to account for the momentum transferred from the radiation to the gas within $r_{\rm trap}$ has previously led to the conclusion that the radiation can never prevent steady-state accretion \citep{begelman1979}.  The main difference between our derivation and this previous work is that we explicitly solve for the condition in which the momentum of the gas exceeds that of the radiation.  Our main new result that follows is that the steady-state assumption underlying the findings in \citet{begelman1979} is only valid when this condition is met.

In other related work, \citet{wyithe12} derive conditions for photon trapping in a model similar to ours, but do not utilize the condition that the momentum of the infalling gas must exceed that of the emitted radiation. 
In another recent analytical study, building on results presented by \citet{inayoshi16}, \citet{sakurai16} utilize 1D models of hyper-Eddington accretion to show that it can occur in instances in which the radiative luminosity generated in the accretion process is sufficiently low, a condition they express in terms of the temperature and density of the ambient medium, the BH mass and the location of the photosphere.\footnote{\citet{sakurai16} also provide an analytical treatment of the dynamics of a shell of accreting material, which is distinct from our assumption of a steady-state flow.}  Though the condition they find for hyper-Eddington accretion is more complex than that for sustained super-Eddington accretion in our Equation~\ref{eq:f_edd_crit}, their finding that hyper-Eddington accretion occurs when the ram pressure of the infalling gas overpowers the radiation which is trapped near the central region around the BH is broadly consistent with our result.

Full 2D \citep{oshuga05} and 3D (\citealt{sadowski16,jiang2019}) simulations of BH accretion have also shown that photon trapping is effective in super-Eddington accretion disks within 100~$R_{\rm S}$, the value of $r_{\rm eq}$ when the radiative efficiency has a canoncial value of $\epsilon$ = 0.1.  Again, this is broadly consistently with our main result that radiation trapping at these scales can facilitate sustained super-Eddington accretion, in particular taking into account the lower radial infall velocity of gas passing through an accretion disk.  That said, one key feature of such accretion calculations not captured in our simple model is that outflows are typically produced around the accretion disk, carrying away of order 10 percent of the mass in the accretion flow (e.g. \citealt{kitaki2021}; but see also \citealt{Hu2022a}), thus somewhat limiting the rate of BH growth though not appreciably slowing infall through the disk itself.\footnote{Outflows driven by precessing BH jets are also not captured in our analytical model, although previous work suggests it is unlikely that precession occurs on timescales shorter than the accretion timescale, on which a super-Eddington inflow is established \citep{kn2013}. } Multi-dimensional calculations have also shown that radiation can escape a super-Eddington disk in the direction perpendicular to the disk. Radiation that escapes in this direction, perpendicular to the accretion flow, will not be available to slow the infall of gas in the disk.  This again implies that $f_{\rm Edd,crit}$, which we have derived under the assumption that there is no disk from which radiation can escape, is an upper limit.

Finally, we note that the results of recent simulations suggest that the gas supply needed for appreciable sustained super-Eddington BH growth may only be available within host galaxies that are themselves rapidly growing.  Simulations of super-Eddington BH growth in isolated galaxies (e.g. \citealt{sassano2022}) and over relatively brief periods following initial BH formation (e.g. \citealt{regan2019}) have shown that feedback effects, including those from supernovae and BH jets, can drive the gas out from the vicinity of the BH, precluding sustained episodes of super-Eddington growth.  Sustained super-Eddington accretion at rates above the limit given by Equation~\ref{eq:f_edd_crit} has, however, been realized in full cosmological simulations in which there is an ever-increasing gas supply in a rapidly growing halo \citep{scoggins2022}.  We may therefore expect that sustained super-Eddington growth of BHs is most likely to be realized in gas-rich galaxies that are rapidly growing in the early universe.

\begin{figure}
\centering
\resizebox{9.3cm}{!}{\includegraphics{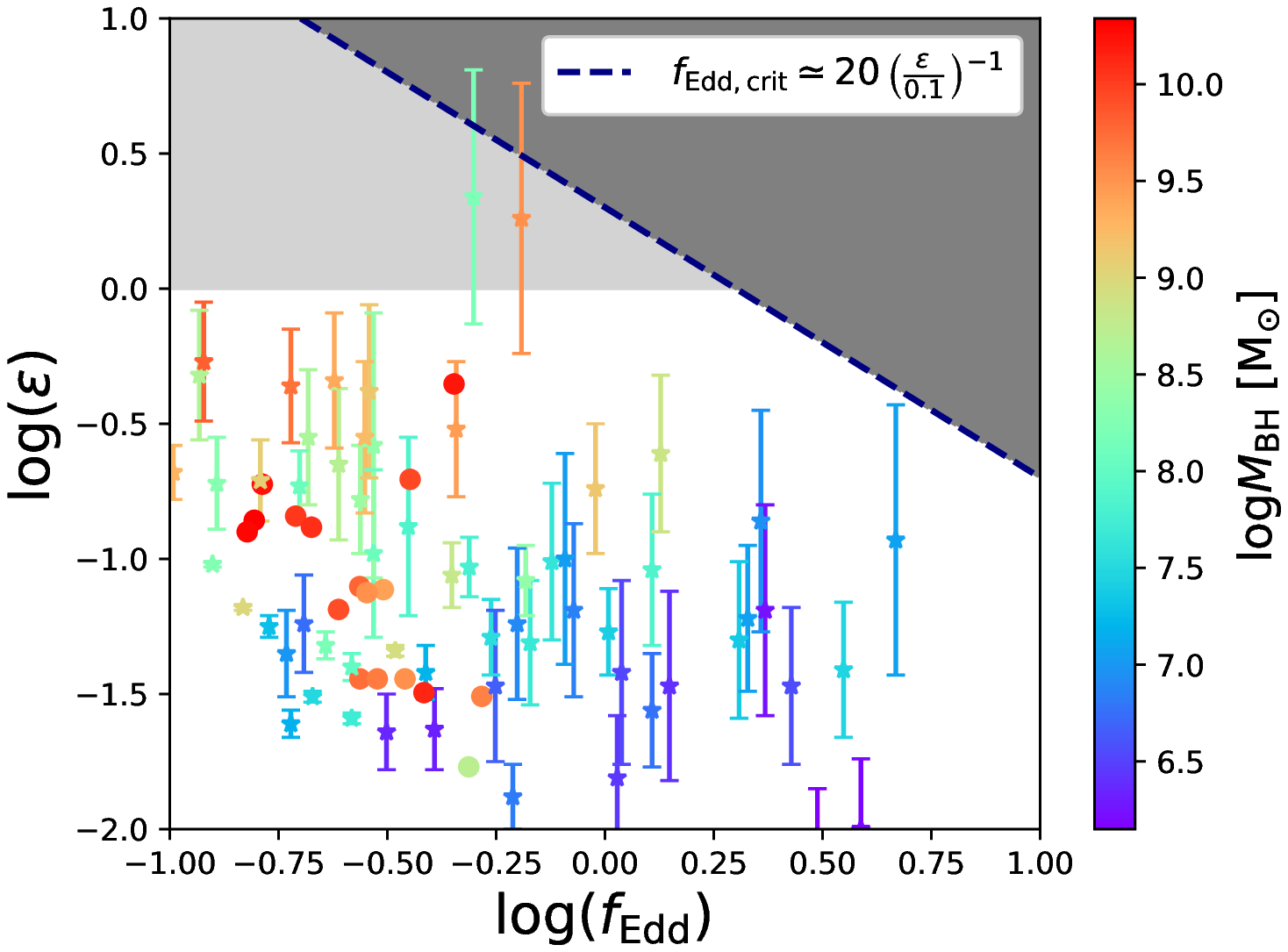}}\\
\caption{Radiative efficiencies $\epsilon$ and Eddington ratios $f_{\rm Edd}$ inferred for BHs powering AGN at $z$ $\ga$ 6 \citep{trakhtenbrot17} and $z$ $\la$ 0.5 \citep{davis11}, with color corresponding to inferred BH mass. Above the dashed line, defined by Equation~\ref{eq:f_edd_crit}, sustained super-Eddington accretion takes place.  As a result of radiation being trapped in the accretion flow, detection of BHs accreting in this mode is expected to be difficult and, indeed, none are identified. The light grey shaded area shows the unphysical regime in which $\epsilon>1$. }
\label{fig:eta_v_f_Edd}
\end{figure}

\section{Comparison to data}
\label{sec:data}
A prediction that follows from Equation~\ref{eq:f_edd_crit} is that black holes powering active galactic nuclei (AGN) that are accreting at greater than $f_{\rm Edd,crit}$ $\simeq$ 20 ($\epsilon$/0.1)$^{-1}$ times their respective Eddington rates may not be detectable due to the radiation emitted in the accretion process being trapped in the accretion flow.  Here we test this simple prediction using the measured values of $\epsilon$ and $f_{\rm Edd}$ for observed AGN.

We use the measurements compiled in \citet{davis11} and \citet{trakhtenbrot17} on BHs powering AGN at $z$ $\la$ 0.5 and $z$ $\ga$ 6, respectively, and directly compare these data to our formula for $f_{\rm Edd,crit}$ given by Equation~\ref{eq:f_edd_crit}, in Figure~\ref{fig:eta_v_f_Edd}.  Along with this comparison, the data points shown in Figure~\ref{fig:eta_v_f_Edd} are color-coded by the masses inferred for each of the BHs.  The majority of the data are well below the limit, with some AGN exhibiting Eddington fractions of a few (see also \citealt{kollmeier06}; \citealt{willott10}; \citealt{wu13}; \citealt{yang21}).  That said, the AGN with the highest measured radiative efficiencies and Eddington fractions lie just below the limit, consistent with the prediction that any AGN above this limit are not detected due to photon trapping.

While none of the AGN shown in Figure~\ref{fig:eta_v_f_Edd}, for which both Eddington fraction and radiative efficiency measurements are available, appear above the line defined by Equation~\ref{eq:f_edd_crit}, future measurements of the radiative efficiencies of AGN with exceptionally high Eddington fractions (e.g. \citealt{tang19,tortosa22}) could potentially challenge the theory. These measurements can be difficult to make \citep{raimundo12}, as photon trapping and obscuration effects potentially lead to underestimates of the efficiency with which radiation is generated in the accretion process near the BH (e.g. \citealt{begelman17,pacucci17,davies19,ishibashi21}) and anisotropic escape of radiation from the accretion flow can be complicating (e.g. \citealt{oshuga05,pognan2020}).\footnote{As noted in \citet{davis11}, the two points with the highest radiative efficiencies shown in Figure 2 may also be erroneous, perhaps due to overestimates of their FUV slopes.}  

The prediction that highly super-Eddington sources would be heavily obscured is bolstered by the expectation that a large supply of dense gas is required to fuel such elevated accretion rates, as this gas would further attenuate any radiation which is not trapped in the accretion flow. Consistent with this, recent observations of highly obscured quasars at $z$ $\ga$ 7 which may be powered by BHs accreting at super-Eddington rates (\citealt{Fujimoto2022}; \citealt{ALMA2022}) strengthen the possibility that upcoming next-generation facilities such as the {\it Rubin Observatory} (\citealt{lsst2019}), {\it Euclid} (\citealt{euclid2019}) and the recently-launched {\it James Webb Space Telescope} (\citealt{gardner2006}) may uncover evidence that the progenitors of the BHs powering such quasars are grown through sustained super-Eddington accretion.

\section{Implications}
\label{sec:implications}
We have shown that BHs can be grown 
in an essentially gas-limited fashion if the critical accretion rate of $f_{\rm Edd,crit}$ $\simeq$ 2/$\epsilon$ is exceeded. This implies that, given a sufficiently large gas supply, a BH accreting in this regime can grow to SMBH scales in well under the time limits
implied by the existence of $z$ $\ga$ 6 quasars hosting $\ga$ $10^9$M$_{\odot}$ supermassive BHs.  Recent cosmological simulations provide support for this possibility, showing that a gas supply sufficient to fuel sustained super-Eddington accretion may allow BHs to grow from 10$^5$ to 10$^9$ M$_{\odot}$ in rapidly assembling galaxies at $z$ $\ga$ 6 \citep{scoggins2022}.  Support is also provided by recent observational results suggesting that the gas density in high-$z$ galaxies increases rapidly with redshift out to $z$ $\ga$ 7.5 \citep{Gilli2022}, implying that there may be an ample supply of dense gas for sustained super-Eddington accretion going back to early epochs of BH growth (see also \citealt{Venemans2017}; \citealt{trinca2022}).


BHs undergoing such sustained super-Eddington accretion may do
so without a radiative signature or with only a weak one, in particular if the radiation emitted in the
accretion process is not able to escape outside $r_{\rm eq}$.  This
implies that there may, in fact, be BHs growing in this manner that
have so far gone undetected.  Indeed, no observed BHs lie beyond the
limit in our Figure~\ref{fig:eta_v_f_Edd}.  If they are unveiled by future observations, this would add to the evidence for  significant obscured BH growth taking place in the early universe (e.g.  \citealt{treister13,comastri15,pezzulli17b,ALMA2022,Fujimoto2022}) and provide strong evidence that sustained super-Eddington growth may explain the emergence of the most massive BHs inferred to power high-$z$ quasars.  
The expectation that highly super-Eddington BHs would be heavily obscured not only due to the trapping of radiation in the accretion flow, but also due to attenuation by the surrounding dense gas from which they are accreting is, in fact, broadly consistent with the recent finding of \citet{Gilli2022} that 80-90 percent of supermassive BHs at $z$ $\ga$ 6 may be hidden from view by the dense gas in their host galaxies. Thus, while direct comparison with the limit shown in Figure~\ref{fig:eta_v_f_Edd} may be observationally challenging, there does exist indirect evidence that the earliest quasars may indeed be powered by BHs that grow via sustained super-Eddington accretion of high density gas.


An additional possible indirect probe of obscured super-Eddington BH growth lies in the size of the photoionized regions surrounding high-$z$ quasars.  The lifetimes for the quasars that are inferred from the size measurements of these H~II regions are typically of the order of 1 Myr (e.g. \citealt{Khrykin2019,morey21,worseck21}) and in some cases only of order 10$^5$ yr (e.g. \citealt{Eilers2017}).  If the bulk of the growth of the BHs powering these quasars occurs in the sustained super-Eddington regime, these observations may be explained since much of the radiation emitted in the accretion process would not have escaped the accretion flow, and that which did would be attentuated by the surrounding dense reservoir of gas, leaving little radiation available to ionize the intergalactic medium.  While this scenario may require extraordinarily large reservoirs of gas to be rapidly supplied to fuel the growth of BHs with masses in the observed range of $\ga$ 10$^9$ M$_{\odot}$, we emphasize that the existence of these BHs at $z$ $\ga$ 6 already implies that such large reservoirs must have been supplied at relatively high rates in order to grow them so quickly after the Big Bang.

In addition to observations that can constrain the role of sustained super-Eddington accretion in the formation of supermassive BHs, cosmological simulations of their growth can, in principle, also provide invaluable insight.  The simple form of our expression for the critical accretion rate above which radiation is trapped in the accretion flow
makes it easy to employ in cosmological simulations or
semi-analytical models of galaxy formation, for instance by simply turning off radiative feedback from an accreting BH when the accretion rate exceeds $f_{\rm Edd,crit}$.  This would allow to assess in
more detail the environments in which sustained super-Eddington
accretion can rapidly grow high-$z$ BHs.  















\section*{Acknowledgements}
Work at LANL was done under the auspices of the NNSA of the US Department of Energy. P.~R.~U.~S. is supported  by a LDRD Director’s Postdoctoral Fellowship. The authors thank Michael Tremmel, Nicole Lloyd-Ronning, Hui Li, Zoltan Haiman, and anonymous reviewers for valuable comments.

\bibliography{References}

\end{document}